\documentstyle[preprint,aps]{revtex}
\begin{document}
\title{Morphological instability of steps during crystal growth from
solution flow\footnote{will be appeared in Journal of Crystal Growth}}
\author{Serge Yu. Potapenko\footnote{E-mail: potap@appl.sci-nnov.ru}}
\date{12-06-95}
\address{Institute of Applied Physics, Russian
Academy of Sciences\\
46 Ul`anov street, 603600 Nizhny Novgorod,
Russian Federation}
\maketitle
\begin{abstract}
It is shown that step moving to meet solution flow can be unstable against
lateral perturbations. The instability of long-wavelength perturbations
occurs at values of the solution flow intensity less than some critical
value depending on the step velocity. At given intensity of the solution
flow, the instability comes at the step velocity exceeding a critical
velocity. Decay of short-wavelength fluctuations is conditioned by the line
tension of the step. The step moving along the solution flow is laterally
stable at all values of the step velocity and the intensity of the
solution flow. The overlapping diffusion field of the neighbour steps
suppresses the lateral instability but it gives an instability of the step
train against doubling of the period, i.e.  neighbouring steps are
attracted.  The equidistant train moving to meet the solution flow is
stable against the period variations.
\end{abstract}

\section{Introduction}
The appearance of numerous investigations of step dynamics on the surface
of growing crystals both experimental and theoretical is evidently
stimulated by two factors.  Firstly, development of technologies to obtain
high quality crystal structures requires deep understanding of elemental
surface processes during the crystal growth. The control of the step
dynamics plays a key role to provide the stable laminar growth, that gives
high quality crystals. The experimental results had been mainly obtained
by ex situ observations and by in situ very sensitive optical phase
contrast or differential interference contrast microscopy.  For the growth
from solutions, observations of ``live'' mono-molecular steps on growing
crystal were performed with the novel in situ optical technique \cite{Ts}.
Secondly, the advent of scanning tunneling and atomic force microscopy has
open the door to investigate atomic scale structure of the crystal
surfaces.  In this measurements, the nanometer-scale precision can be
achieved not only in the perpendicular to the surface direction as for the
optical observations, but also along the surface. The direct observation
during crystal growth from solution has been realized \cite{Hansma}.

Most studies have been devoted to the growth from vapour phases. Barton,
Cabrera and Frank (BCF) in their seminal paper \cite{BCF} who considered a
step flow model, outlined this case. A vicinal surface consists of broad
terraces separated by monoatomic steps. The transfer of the crystallizing
matter includes  adsorption upon the solid, diffusing across the terraces
during some lifetime and further desorption back into the vapour phase or
attachment to a step.\footnote{One dimensional diffusion along the step is
also possible while the adatoms build into the solid} BCF considered the
steps as a perfect sink, when the attachment kinetics was neglected and
the surface concentration of the substance had a local equilibrium value,
taking into account the curvature correction via the Gibbs-Thomson effect.
Taking into account of the kinetics of the adatoms attachment introduces,
in general, two kinetic coefficients corresponding to adatom attachment
from the lower and upper terraces \cite{Schwoebel,Bennema}. A linear
stability analysis within quasi-static approximation \cite{Bales} predicted
that the step can undergo a morphological instability at a step velocity
of more than a critical value, only if the attachment rate from lower
terrace exceeds the upper one. The instability mechanism is an analog of
the Mullins-Sekerka instability \cite{MS}. More general studies of the
stability both the lateral and the longitudinal has been recently
performed \cite{Liu,Saito,Pimpinelli} without using a quasi-static
approximation. It was shown that the lateral instability can still take
place for symmetric step kinetics because of an asymmetry in forward and
backward direction of the step since it is in motion.

The papers cited above are devoted to the linear stability analysis
against small fluctuations. Scenari\`o of further evolution of the step
pattern was studied in Ref. \cite{Bena}.  This is evidence of presence
of the spatiotemporal chaos in the step dynamics of the BCF model at the
asymmetrical attachment kinetics. A competition between the thermal noise
and the determinism was addressed in Ref. \cite{Karma}.

Step bunching can be induced by an electromigration force during
sublimation of vicinal surfaces, when the heating source is an electric
current \cite{Stoyanov,Natori,Houchman}. Adatoms on the surface are
assumed to carry electric charges and therefore their Brownian motion is
modified by the electric field. The diffusion equation for adatom
concentration $c$ contains the term proportional ${\bf F\nabla} c$, where
$F$ is the electric force. The presence of the step bunching depends both
on the temperature and on the direction of the heating current. In
particular, reversing the current direction transforms a stable
temperature region into an unstable one and vice versa.

In contrast with the growth from gas phase, there are no convicting
evidences on the mechanism of the substance transfer during growth from
solutions. At the solution growth, the solute concentration is closer than
the vapour density to the crystal density and the solution diffusivity is
scarcely less than the surface diffusivity. Therefore, the volume
diffusion in the solution with a direct incorporation into the step can be
basic process transferring the crystallizing units \cite{ChUsp}. On the
other hand, the direct incorporation can be suppressed by an interaction
of the units with the solution and there is a possibility of the
competition between the volume diffusion and the surface diffusion. In
Ref.  \cite{Hansma1}, in situ observation of the dynamics of monomolecular
growth step on the $(10\overline{1}4)$ cleavage surface of calcite was
presented and it was concluded that the surface diffusion does not control
the calcite growth.  However, the kinetic coefficient of these steps is
very small $\sim 2\times 10^{-7}$ cm/c, and the step motion is not
controlled by any diffusion. The only attachment kinetics determines the
step velocity. At the same time, there are reasons for the surface
diffusion at growth of the KDP crystals \cite{Vekilov,DeYoreo}. We believe
that investigations of step dynamics peculiarities at the volume diffusion
can flow a light on this question.

A formation of step bunches or a longitudinal instability forced by the
influence of solution flow within the diffusion boundary layer was
observed in Ref. \cite{Ch1} and has been considered theoretically in Ref.
\cite{Ch}.  The stepped vicinal was described as a continuous surface with
kinematic waves of the step density. It was found that the step bunching
depends on the angle of the riser and way of the heat transfer in the case
of the melt growth. It can be developed only at the parallel motion of the
straight steps and the solution.  In Ref.  \cite{train}, the kinetics of
the rectilinear macrosteps under diffusion or thermal interactions in
stagnant media has been studied when the velocities of the steps are
controlled by the supersaturation at the feet of the steps.
The stability of the macrostep train depends on the angle of the step
risers and way of the heat transfer in the case of the melt growth.

The present study addresses the zig-zag instability of steps and the
instability of the step train against doubling of the period during the
crystal growth in the solution flow.

\section{Green's function for convective diffusion}
Let us consider the diffusion nearby the vicinal crystal face being in a
steady solution flow.  We suppose that the height of the step shown in
Fig.  1 is small in compare with the thickness of the diffusion layer and
the characteristic curvature radius of the step, so the steps can be
considered as a linear sinks for the diffusion field on the flat crystal
surface.  At large values of the Prandtl number $Pr=\nu/D\gg 1$, where
$\nu$ is the kinematic viscosity and $D$ stands for diffusivity, the
thickness of the diffusion layer is much less than the thickness of the
hydrodynamic boundary layer.  The solution velocity can be developed in a
series in perpendicular to the surface direction $y$.  The solution
velocity at the crystal surface vanishes and the velocity component
parallel to the surface in the first order is
\begin{equation}
v_z=B(z) y.
\end{equation}
The perpendicular component $v_y$, being determined by equation of
continuity
\begin{equation}
{\partial v_y\over\partial y}+{\partial v_z\over\partial z}=0,
\end{equation}
has the form
\begin{equation} v_y=-{y^2\over 2}{d B\over d z}.
\end{equation}
The component $v_y$ is in inverse proportion to a characteristic length of
variation of the solution flow along the surface $L_{flow}$, i.e. in fact
the size of the crystal face. Since the diffusion fields, varying in
sufficient less scales, are of our interest, so the terms with $v_y$ can
be omitted in diffusion equation and the parameter of the solution
intensity $B$ can be taken independent on $z$ within the given part of the
crystal surface.  Then the diffusion equation for the solute concentration
$C$ takes the form
\begin{equation}
{\partial C\over\partial t}+B y {\partial
C\over\partial z} = D\left({\partial^2 C\over\partial x^2}+{\partial^2
C\over\partial y^2} +{\partial^2 C\over\partial z^2}\right) - j(x,z)
\delta(y),
\label{dif}
\end{equation}
where $j(x,z)$ is the substance flux crystallized in this point of the
surface. Let the solution velocity $v_z>0$ and so $B>0$, whereas the step
velocity can be both positive and negative. After the Fourier
transformation from $t$, $x$ and $z$ to $\omega$, $p_x$ and $p_z$ we
obtain the equation and the boundary condition
\begin{equation}
{d^2 C\over d\xi^2} - (\xi_0+\xi) C=0;
\;\;\; \left.{d C\over d\xi}\right|_{\xi=0}
=(B D^2 i p_z)^{-1/3} j,
\label{dify}
\end{equation}
where $\xi = (i p_z\Lambda^{-2})^{1/3} y$ and $\xi_0 =(i
p_z\Lambda^{-2})^{-2/3} (i \omega/D +p_x^2+p_z^2)$, $\Lambda^2=D/B$.
The solution of Eq. (\ref{dify}) is expressed through the Airy function
vanished at $y\rightarrow\infty$:
\begin{equation}
C=j(B D^2 i p_z)^{-1/3}
\hbox{Ai}(\xi+\xi_0)/\hbox{Ai}'(\xi_0).
\end{equation}
Here at $p_z>0$, in order to satisfy the boundary condition at infinity we
have to suppose $i p_z=p_z\exp(i\pi/2)$ and at $p_z<0$:  $i
p_z=|p_z|\exp(-i\pi/2)$.  Taking these conditions into account, we obtain
the Green's function of Eq.  (\ref{dif}) in the form
\begin{eqnarray}
G(x,y,z,t)&=&{2\over (2\pi)^3} \int_{-\infty}^{\infty}dt\;
\int_{-\infty}^{\infty}dp_x\;\int_0^{\infty}dp_z\;
\hbox{e}^{i\omega t +i p_x x}
\hbox{Re}\left\{\hbox{e}^{i p_z z}
\left ({\Lambda^2\over i p_z}\right )^{1/3}
\times\nonumber\right.\\
&&\left.{\hbox{Ai}\left [\left (i p_z\Lambda^{-2}\right )^{1/3} y+
(i p_z\Lambda^{-2})^{-2/3}
(i\omega D^{-1}+p_x^2+p_z^2)\right ]\over
\hbox{Ai}'\left [\left (i p_z\Lambda^{-2}\right )^{-2/3}
(i\omega D^{-1}+p_x^2+p_z^2)\right ]}\right\}.
\label{Green}
\end{eqnarray}

Consider the diffusion field formed by the rectilinear steps moving as $z=V
t$ at the constant solute flux $j$ per unit of the step length. If $C_0$
is the solution concentration of the incoming flow, then
\begin{eqnarray}
ó_0-C(y,z-Vt)&=&-{j\over \pi D}
\int_0^{\infty}dp_z\;
\hbox{Re}\left\{\hbox{e}^{i p_z (z-Vt)}
\left ({\Lambda^2\over i p_z}\right )^{1/3}
\times\nonumber\right.\\
&&\left.{\hbox{Ai}\left [\left (i p_z\Lambda^{-2}\right )^{1/3} y+
(i p_z\Lambda^{-2})^{-2/3}
(p_z^2-V D^{-1} i p_z)\right ]\over
\hbox{Ai}'\left [\left (i p_z\Lambda^{-2}\right )^{-2/3}
(p_z^2-V D^{-1} i p_z)\right ]}\right\}.
\label{line}
\end{eqnarray}
Nearby the step, the concentration is logarithmical dependent on the
distance to the step $r=\sqrt{y^2+\zeta^2}$, $\zeta=z-Vt$. The main
contribution in the integral (\ref{line}) goes from the large values of
$s=\Lambda p_z$, when the ratio of the Airy function to its derivative
tends as
\begin{equation}
F(s)={\hbox{Ai}\left [(i s)^{1/3} y/\Lambda+(i s)^{-2/3}
(s^2-\Lambda^2 V D^{-1} i s)\right ]\over
\hbox{Ai}'\left [(i s)^{-2/3}
(s^2-\Lambda^2 V D^{-1} i s)\right ]} \rightarrow
-s^{-2/3}\hbox{e}^{i\pi/6-s y/\Lambda}.
\label{as}
\end{equation}
To use this asymptote as an approximation we have to correct the factor
before exponent to provide the right value at $s=0$:  $F_0=F(0)=-3^{-1/3}
\Gamma(1/3)/\Gamma(2/3)$, where $\Gamma(z)$ is the gamma function. $F(s)$
can be approximated by the function
\begin{equation}
F(s)=-(s+s_0)^{-2/3}\hbox{e}^{i\pi/6-s y/\Lambda},\;\;
s_0= \hbox{e}^{i\pi/4} |F_0|^{-3/2},
\label{approxF}
\end{equation}
for which integrating in (\ref{line}) gives
\begin{equation}
C_0-C={j \over\pi D}\Gamma(2/3)
\hbox{Re}\left\{\Psi\left[2/3,1;s_0\Lambda^{-1}(y-i\zeta)\right]\right\}.
\label{hyper}
\end{equation}
Here $\Psi(a,b;z)$ is the confluent hypergeometric function \cite{BE}. In
the limit $r\rightarrow 0$, we have
\begin{equation}
C_0-C(r)={j\over\pi D}\ln{\alpha_0\Lambda\over r},\;\;\;\;
\alpha_0=3^{-1/2}
\left[{\Gamma(1/3)\over\Gamma(2/3)}\right]^{3/2}
\hbox{e}^{2\psi(1)-\psi(2/3)}
\approx 1.90,
\label{ln}
\end{equation}
where $\psi(z)=\Gamma'(z)/\Gamma(z)$ is the digamma function. The
numerical integration gives $\alpha_0\approx 2.81$.

If the step height is small, $h\ll\Lambda$, then the step velocity is
controlled by the concentration (\ref{ln}), where one has to put
$r\approx h$. For the linear step kinetics,
\begin{equation}
V=\beta\left[C_e-C(h)\right]/C_e,
\label{k}
\end{equation}
where $\beta$ is the kinetic coefficient and $C_e$ is the concentration of
the saturated solution, taking into account $j=\rho h V$, we obtain
\begin{equation}
V={\beta \sigma_b\over
1+\rho\beta h\, (\pi D C_e)^{-1}\;\ln(\alpha_0\Lambda /h)}.
\label{V}
\end{equation}
Here $\sigma_b =(C_0-C_e)/C_e$ is the bulk supersaturation and $\rho$ is
the crystal density.

Let us investigate the asymptotes of the concentration field far from the
step within quasi-static approximation. At $y=0$ and $\zeta>0$ in the limit
$\zeta\rightarrow\infty$, the expression (\ref{hyper})  is exact because,
in this case, the main contribution of the integral (\ref{line}) goes from
the small values of $p_z$, where the approximation (\ref{approxF}) is
exact.  At $\zeta\gg\Lambda$, using asymptotic development of the
confluent hypergeometric function \cite{BE} we arrive to the expression
for the surface concentration
\begin{equation}
C_0-C(\zeta)=
j\;{3^{1/6} \Gamma(1/3)\over 2\pi D}
\left({\Lambda\over \zeta}\right)^{2/3},\;\;\;\;  \zeta>0.
\label{z>0}
\end{equation}
To find the surface concentration at the side of flow incoming in the
limit $\zeta\rightarrow -\infty$, we change the integration variable
in (\ref{line}) to $\tau=i p_z\Lambda$:
\begin{equation}
C_0-C(\zeta)={j\over\pi D}
\hbox{Im}\int_0^{i\infty} d\tau \tau^{-1/3} \hbox{e}^{\zeta\tau/\Lambda}
{\hbox{Ai}(-\tau^{4/3})\over\hbox{Ai}'(-\tau^{4/3})}.
\label{z<0-1}
\end{equation}
At $\zeta<0$, the path of the integration in the $\tau$-plain shown in
Fig.  2 can be reduced to the positive part of the real axis. The
integrable function has poles only on this new path, where the function is
real.  So the imaginary part of the integral is the half-sum of residuals
in the poles. If $|\zeta|\gg\Lambda$, the main contribution goes from the
closest to zero pole $\tau=(a_1')^{3/4}$, where $a_1'\approx 1.01879$ is
the least root of the equation $\hbox{Ai}'(-a') =0$. The result is
\begin{equation}
C_0-C(\zeta)={3j\over 4D} (a_1')^{-3/2}
\exp\left[-(a_1')^{3/4}
\left|{\zeta\over\Lambda}\right|\right],\;\;\;\;\zeta<0.
\label{z<0}
\end{equation}
The concentration field relaxes slow along the solution flow and fast in
the opposite direction. Analogous calculation for the concentration far
from the crystal surface leads to the expression
\begin{equation}
C_0-C(y)={3j\over 8\pi^{1/2} D\; (a_1')^{25/16}
\hbox{Ai}(-a_1')} \left({y\over\Lambda}\right)^{9/4}
\;\exp\left[-{2\over 3} (a_1')^{3/8}
\left({y\over\Lambda}\right)^{3/2}\right].
\label{Ch}
\end{equation}
Note that the quantity $\Lambda$ can be interpreted as an effective
thickness of the diffusion layer in this case.

At the large step velosities, $V\sim D\Lambda^{-4/3} \zeta^{1/3}$, the
quasi-static approach is not correct.
In the limits $V\gg D\Lambda^{-4/3} \zeta^{1/3}$ and
$|\zeta|\rightarrow\infty$, the diffusion field reads as
\begin{equation}
C_0-C(\zeta)= j\times\left\{\begin{array}[c]{ll} \left(\pi
V D \zeta\right)^{-1/2},
& \zeta>0\\ 3 D^{-1} \exp\left[-(a_1')^{1/3}
w^{-3} \Lambda^{-1}|\zeta|\right], & \zeta<0 \end{array}\right.,
\label{nonst}
\end{equation}
where $w=V\Lambda/D$ characterizes the degree of the quasi-stationarity of
the diffusion. Like the quasi-static case, the zone impoverished by the
step sink is drifted by the solution flow. The results obtained in this
Section will be used for analysis of the step train stability.

\section{Stability of one isolated step}
A step can be treated as isolated if the diffusion length $\Lambda$ is
small in compare with the step spacing. Certainly, the isolated step
can contain several elementary crystal layers. We shall assume
that scale of internal structure, such as height and width of this
composite step is the least of sufficient length parameters. For the
isolated step, the scale plays a role of effective step height $h$ which
determines the concentration (\ref{k}) controlled the step motion.

Near by the crystal surface, the concentration up the solution stream is
higher than the concentration down the stream. The isolines of the
concentration field are schematically shown in Fig. 3.
It is clear that a
negative local deviation of the step from a straight, $\delta z(x)<0$,
being directed upstream, finds the more concentrated solute and the
opposite deviation meets the impoverished solution. Further evolution of
the perturbation depends on the relation between directions of the step
motion and the solution flow.  For the step moving along the flow, being
at the less concentration the leading part of the step becomes slower.  At
that time, finding the larger concentration the dropping behind part
overtakes the rest step.  Therefore, the lateral perturbations of the step
moving along the flow always decay.  At the case of the opposite motion,
the velocity increment of the perturbed part is having the same sign with
the variation of the z-coordinate of the step.  The presence of this
positive back-coupling can produce a spontaneous growth of the
perturbation, i.e.  to the instability of the straight step shape. The
line strain of the step and forcing the solution flow stabilize the
straight shape.

Thus, the flow of the solution gives rise an analogous effect to the
Schwoebel's barrier \cite{Schwoebel} at the surface diffusion transport.
In the case of the volume diffusion, the substance flux is larger from the
side of the solution incoming than from the opposite side. At the surface
diffusion, the fluxes arriving from the opposite terraces are differ due
to a difference of the energy barrier of the adatom attachment from the
upper and the lower terraces.  The feature of the volume diffusion is in a
possibility to govern the quantity of the attachment asymmetry by the
direction and the intensity of the solution flow.

Consider the kinetics of small fluctuations, $z=V t+\delta z(x,t)$, from
the straight step. The unperturbed step velocity $V$ is determined by
(\ref{k}). Taking into account the Gibbs-Thomson effect, the kinematics
equation takes the form
\begin{equation}
\dot z= \pm\beta\tilde\sigma (z,t)
+\beta\lambda  z''
\left[1+(z')^2\right]^{-3/2},
\label{kin}
\end{equation}
where $\rho_c(\tilde\sigma)=\lambda/\tilde\sigma$ is the critical nucleus
radius on the crystal surface at the supersaturation $\tilde\sigma$ near
by the given point of the surface and $\lambda$ is the capillary length.
The upper sign (\ref{kin}) corresponds to the parallel motion of the step
and the flow and the lower sign corresponds to the opposite motion.
Assuming $\delta z(x,t)=\epsilon(t) \cos(qx)$, let us calculate the
concentration near by the step at $\epsilon q\ll 1$. The substance flux
per the unit length is $j=\rho h \left[V+\dot\epsilon\cos(qx)\right]$.
Using the Green's function (\ref{Green}), we obtain
\begin{equation}
\tilde\sigma (x,t)= \sigma_b - (\pi U)^{-1} V
\,\ln(\alpha_0 /H)-
\left[\dot\epsilon I_0(H,Q)
+ \epsilon {V\Lambda^{-1} I(Q)}\right] U^{-1} \cos(qx),
\label{sigma}
\end{equation}
\begin{equation}
I_0(H,Q)= -{1\over\pi}\,\hbox{Re}\int_0^{\infty}
{d\tau\over (i\tau)^{1/3}}\;
{\hbox{Ai}\left[H (i\tau)^{1/3} + (i\tau)^{-2/3}(\tau^2+Q^2)\right]
\over\hbox{Ai}'\left[(i\tau)^{-2/3}(\tau^2+Q^2)\right]},
\label{I0}
\end{equation}
\begin{equation}
I(Q)= {1\over\pi}\,\hbox{Re}\int_0^{\infty}
d\tau\; (i\tau)^{2/3}\;
\left\{
{\hbox{Ai}\left[-(i\tau)^{4/3}\right]
\over \hbox{Ai}'\left[-(i\tau)^{4/3}\right]}
-{\hbox{Ai}\left[(i\tau)^{-2/3}(\tau^2+Q^2)\right]
\over\hbox{Ai}'\left[(i\tau)^{-2/3}(\tau^2+Q^2)\right]}
\right\},
\label{I}
\end{equation}
where $U=D C_e/(\rho h)$, $H=h/\Lambda$ and $Q=q\Lambda$.
Taking into account (\ref{sigma}) reduces the kinematic equation (\ref{kin})
to
\begin{equation}
\dot\epsilon=\nu^{\pm}\epsilon,\;\;\;\;
\nu^{\pm}={\beta\over\Lambda}\,\cdot {\mp V U^{-1} I
- \lambda\Lambda^{-1} Q^2
\over 1+ \beta U^{-1} I_0},
\label{nu}
\end{equation}
where $\nu^{\pm}$ are the amplification rates for the parallel and the
opposite motion respectively. At positive $\nu$, the perturbation amplitude
$\epsilon$ is exponential growing in time and negative values imply
decaying.  The function $I_0(H,Q)$ and $I(Q)$ are always nonnegative.  The
dependences $I_0$ on $H$ at several values $Q$ are shown in Fig. 4.  At
$H\ll 1$, they can be written as
\begin{equation}
I_0(H,Q)=\pi^{-1} \ln\left[\alpha(Q)/H\right].
\label{alpha}
\end{equation}
The dependence $\alpha^{-1}(Q)$ is shown in Fig. 5, where
$\alpha(0)=\alpha_0$, $\alpha(\infty)=2\,\exp(-\gamma)$ and
$\gamma\approx 0.5772$ is the Euler number.
The dependence $I(Q)$ are shown in Fig. 6. At $Q\ll 1$, we have
$I(Q)\approx 0.19 Q^2$ and $I(\infty)\approx 0.128$. With a relative
error below 1\% the functions $I_0$ and $I$ can be approximated by the
expressions
\begin{eqnarray}
I_0(H,Q)&=&(2\pi)^{-1} \ln\left[(0.77+0.12 Q^2)/H\right],\\
I(Q)&=&0.19 Q^2/(1+1.58 Q^2).
\label{approxI}
\end{eqnarray}
The expression (\ref{nu}) shows that the step moving along the stream is
stable for the lateral disturbances at all wavelengths. At the opposite
motion, if $I''(0)>U\lambda/(V\Lambda)$, then the step is unstable for the
long wave disturbances with $q<q_{cr}$, where
$q_{cr}\approx\Lambda^{-1} [0.12 \Lambda V/(\lambda U)-0.63]^{1/2}$. The
instability condition is given by
\begin{equation}
{B^{1/2}\over V h}< {0.19 \rho\over C_e \lambda D^{1/2}}.
\label{cond}
\end{equation}
Thus, the lateral instability occurs at low intensity of the solution flow
or/and at  sufficient high values of the step velocity and the step height.
The instability zone, $0<q<q_{cr}$, of the disturbance wave vectors is
shown by shading in Fig. 7. The thick line, $q_{cr}$, starts at origin
because the step is stable without flow in the frame of quasi-statics. The
minimal wavelength of the unstable disturbances $l^{min}=82.7
(U/V)\lambda$ is determined by the capillary length times the ratio of the
parameter of the diffusion velocity to the step velocity.  Remember that
the quasi-static approximation is correct at $V\Lambda/D\ll 1$.

At low intensities of the solution flow, $B\rightarrow 0$, the
characterictic size of the impoverished zone $\Lambda\rightarrow\infty$,
therefore the instability vanishes in this limit. At high flow
intensities, the concentration near by the step is close to the bulk
concentration at all points of the step that explains for the stability at
large $B$.  Thus the instability vanishes at $q=0$ and $q=q_{cr}$ and the
amplification rate of the instability has a maximum within the interval
$0<q_0<q_{cr}$. If at least one of the inequalities $q_0\Lambda\ll 1$ and
$\beta/(2\pi U)\ll 1$ is fulfilled then
\begin{equation}
q_0\Lambda=0.80\left[\left({0.19\Lambda
V\over\lambda U}\right)^{1/2} -1\right]^{1/2},\;\;\;\;
\nu^-_{max}={0.12\beta V\over\Lambda U}\left[\left({\lambda U\over 0.19
\Lambda V}\right)^{1/2}-1\right]^2.
\label{numax}
\end{equation}
The instability requires the step track length $L=V/\nu_{max}$ to manifest
itself.

The single step approximation addressed in this Section becomes
inapplicable when the step spacing $d$ is comparable to the diffusion
length $\Lambda$.  If a screw dislocation, having the net number $m$ of
unit steps in the normal component of the Burgers vector, generates a step
train, each step consisting of $n$ of unit steps, then
\begin{equation}
d\approx 20 n \rho_c/m.
\label{d}
\end{equation}
At $d<\Lambda$, the single step instability can be realized by a sharp
dropping of the solution temperature from a value at which $d^*\gg\Lambda$
in the steady state. The spacing $d^*$ will have been conserved while new
generated steps have not arrived.

\section{Lateral stability of vicinal face}
To understand the influence of the diffusion interaction of neighbour
steps it is appropriate to consider the inverse limit to the single step
case. At $d\ll \Lambda$, the crystal relief is described by a height
function $h(x,z)$. Volume conservation implies that a continuity
equation satisfies:
\begin{equation}
{\partial{\bf
P}\over\partial t}+{\bf\nabla} ({\bf P V})=0, \;\;\;\;{\bf P}={\bf\nabla}
h, \;\;\;\;
{\bf\nabla}=({\partial\over\partial
x},\,{\partial\over\partial z}),
\label{kinematics}
\end{equation}
where $|\bf P|$ is the surface slope to a singular face and the direction
$\bf P$ is parallel to the step velocity $\bf V$. The normal growth rate
of the crystal face $R={\bf P V}$ and the flux per unit area is $J=\rho
R$.  Using the Green's function (\ref{Green}), for the small fluctuations
from the plain ${\bf P}_0=(0,\,P_0)$ of the form
\begin{equation}
{\bf P}={\bf P}_0+\tilde{\bf P}
\exp(i\omega t+i qx+\kappa z)
\label{perturbP}
\end{equation}
\begin{equation}
R=R_0+\tilde{R}\exp(i\omega t+i qx+\kappa z),
\;\;\;\; R_0={\bf P}_0 {\bf V}_0,
\label{perturb}
\end{equation}
where $R_0$ and ${\bf V}_0= (0,\,V_0)$ are unperturbed normal growth rate
and step velocity, we obtain the surface concentration in the form
\begin{eqnarray}
C&=&C_b-\rho\Lambda D^{-1} T(\omega,\,q,\,\kappa) \tilde{R}
\exp(i\omega t+i qx+\kappa z),
\label{T1}\\
T(\omega,\,q,\,\kappa)&=&
(\Lambda \kappa)^{-1/3}
{\hbox{Ai}\left[
(\Lambda^{-2}\kappa)^{-2/3}
(i\omega D^{-1}+q^2-\kappa^2)\right]\over
\hbox{Ai}'\left[\left(\Lambda^{-2}\kappa\right)^{-2/3}
(i\omega D^{-1}+q^2-\kappa^2)\right]}.
\label{T2}
\end{eqnarray}
These perturbations are the quite lateral only at $\kappa=0$.
The stability against the quite longitudinal perturbations, $q=0$, has
been considered in Ref. \cite{Ch}.  Using the kinematic equation
(\ref{kinematics}), we obtain the linear stability equation
\begin{equation}
i\omega + V_0 \kappa + \rho_c R_0 q^2 -
i T(\omega,\,q,\,\kappa) \omega\rho P_0 \beta \Lambda (D C_e)^{-1} =0.
\label{dis}
\end{equation}
In the limit $\kappa''\ll\kappa'$, where $\kappa'$ and $\kappa''$ are the
real and imaginary parts of $\kappa=\kappa'+i\kappa''$, the linear
dispersion relations are given by
\begin{equation}
\kappa''=-\omega/V_0\equiv \Lambda^{-1} \theta,\;\;\;\;
\kappa'=P_0 \rho C_e^{-1}\beta\theta\, \hbox{Im}(T)-P_0\rho_c q^2.
\end{equation}
The amplification or decay lengths for the step motion up and down the
solution flow (upper and lower sign respectively) are given by
\begin{equation}
\left(L^{\pm}\right)^{-1}= \pm{|P_0|\rho\beta\over C_e D} \theta^{2/3}
\hbox{Im}\left\{ i^{-1/3}\cdot
{\hbox{Ai}\left [(i\theta)^{-2/3}
(Q^2+\theta^2 \mp i w\theta)\right ]\over
\hbox{Ai}'\left [(i\theta)^{-2/3}
(Q^2+\theta^2 \mp i w\theta)\right ]}\right\}
-{|P_0|\rho_c Q^2\over\Lambda^2},
\label{L1}
\end{equation}
where $w=|V_0|\Lambda/D$ takes into account the influence of the step
motion on the diffusion field.  Positive and negative signs of $L$ imply
the instability and the stability respectively. For quite longitudinal
displacements of steps from steady state positions, $Q=0$, the expression
(\ref{L1}) is a particular case of the results of Ref. \cite{Ch}.

Quite lateral fluctuations, $\theta=0$, are always decay,
$L^{\pm}=-|P_0|\rho_c q^2<0$, the decay rate being determined only the
step line tension. The explanation lies in the variation of the
concentration along curved step being compensated with the diffusion
influence of neighbour steps.

It can be numerically shown that at the opposite motion of the steps and
the solution flow, at all values of the parameters, the length $L^-<0$,
hence the strong diffusion interaction completely suppresses the lateral
instability. In the case of parallel motion, there is bunching with
simultaneous amplification of the perturbation along the steps for only
sufficient long-wave perturbations: $\theta<\theta_{cr}(w,Q)$. The
presence of the lateral perturbation, $Q\not=0$, decreases the instability
of the density waves of steps i.e. straight steps are more unstable.  Note
that the diffusion nonstationarity has a stabilizing character: the
instability drops with increasing $w$.

Thus, the diffusion interaction suppresses the lateral instability at
$\Lambda>A\,d$, where $A\sim 1$. For the dislocation steady growth
(\ref{d}), it confines the unstable zone of supersaturation from above. As
a whole, the unstable zone reads as
\begin{equation}
{5.3 \lambda U_0\over \beta n}\left({B\over D}\right)^{1/2}<
\sigma<{20 A \lambda n\over m}\left({B\over D}\right)^{1/2},
\label{<s<}
\end{equation}
where $U_0=D C_e/(\rho h_0)$ and $h_0$ is the height of unit step. This
zone exists only at $n^2/m>0.26 U_0/(A\beta)$.

For an example, evaluate the instability of step on the (101) face of the
KDP crystal. The value of the solution flow intensity is given by
\begin{equation}
B=\left.{\partial^2\psi\over\partial y^2}\right|_{y=0},
\label{B}
\end{equation}
where $\psi$ is the stream function. $B$ can approximately be
evaluated as $B\sim U_{sol}/\delta$, where $U_{sol}$ characterizes the
flow velocity far from the surface and $\delta$ is the thickness of the
hydrodynamic boundary layer at this point. For concreteness, let us take a
tangential flow of solution. Other useful plane mass transfer models have
been described in Ref. \cite{P}.  For the flow around a semi-infinite
plate \cite{Schlichting}, $z>0$, at an incoming velocity $U_{sol}$, we
have $B=0.33(U_{sol}^3/\nu z)^{1/2}$. For $U_{sol}=10$ cm/c and $z=1$ cm
we obtain $B=100$ c$^{-1}$, and the instability zone is given by
$0.3/n<\sigma<0.003 n/m$, it existing if $n^2/m>100$. Taking into account
an anisotropy of the step kinetic coefficient and the isotropy of the
critical nucleus \cite{DeYoreo} weakens this restriction, because the step
spacing in the direction of the largest kinetic coefficient exceeds
$20\rho_c$ \cite{anis}.  At the case of $m=1$, the unit steps are stable
against the lateral fluctuation but the steps containing ten or more units
are unstable.  At $\sigma=0.05$, for the step of 20 unit layers, the
amplification length $L^-\approx 0.15$ cm that can be observed.

\section{Instability of step train at weak interaction}
The diffusion interaction between steps of a train is weak when the step
spacing is considerably greater than the diffusion length: $d\gg\Lambda$.
If the steps are moving to meet the solution flow then the condition of
the lateral instability is given by (\ref{cond}) with a weak stabilizing
influence of the interaction. A sufficient effect of the weak coupling
occurs at the parallel motion of the step and the solution. In this case,
the lateral instability does not develop, so we will assume that the steps
are straights with the coordinates $z_n$. Since the diffusion field going
from a step to meet the solution flow decays faster than along the flow,
we will take into account only influence of the closest step from the
side of solution incoming. The velocity of $n$th step is given by
\begin{equation}
V_n=V-{3^{1/6} \Gamma(1/3)\,\beta\over2\pi D C_e}
\left({\Lambda\over z_n-z_{n-1}}\right)^{2/3}
j_{n-1},
\label{Vn}
\end{equation}
where the unperturbed step velocity $V$ is determined by (\ref{V}).  The
perturbation doubling the train period, $z_n=nd+(-1)^n \tilde{\epsilon}
\exp(\nu t)$, has a maximal amplification rate. The amplification length
for parallel motion and the decay length for antiparallel
are respectively given by
\begin{equation}
L^{\pm}=\pm{3^{5/6} \pi U d\over 2\Gamma(1/3) \beta}
\left(1+{\beta\over\pi U}\, \ln{\alpha_0\over H}\right)
\left({d\over\Lambda}\right)^{2/3}.
\label{Ltrain}
\end{equation}
The magnitude of $\chi=(\beta/\pi U)\ln(\alpha_0/H)$ is a criterion of the
regime of the step motion.  In the limit of the small values, the step
velocity is controlled by the crystallization kinetics, i.e. the kinetic
regime takes place, and at the limit $\chi\gg 1$  the diffusion regime
occurs. Note that even near by the kinetic regime, when the step velocity
practically does not depend on the flow velocity, the instability
nevertheless exists though the amplification rate is small with respect to
this parameter.

The reason of the instability of the train period at parallel motion with
respect to the flow is quite clear because if a step displaces forward
then the step velocity takes an increment due to weakening the diffusion
overlap of the previous step, and at the same time, the next step becomes
slowing down, hence the perturbation will grow. This effect is an
analogous to the current-induced step bunching during sublimation
\cite{Stoyanov,Natori,Houchman} for the step up direction of the force if
the surface diffusion length $\lambda_{2d}$ is small compared to the step
spacing $d$. In this case, the induced migration is directed along the step
motion.  However, the migration force can also induce the bunching for the
step down direction if there is the opposite correlation between the
surface diffusion length and the step spacing. This correlation depends on
the temperature.  Increasing the temperature, the surface diffusion
length decreases exponentially and may become smaller then $d$.  At the
volume diffusion growth and dissolution, the bunching can takes place only
for the parallel motion of the steps and the solution flow at any relation
between the diffusion length $\Lambda$ and $d$. At the surface diffusion,
if the $d$ is small compared to $\lambda_{2d}$, the most adatoms departed
from a step reach the adjacent step to which the migration is directed.
The adatom concentration grows along the force because they can only
incorporate into the step or evaporate. It is precisely this fact that
gives the instability for the opposite direction of the step motion and
the adatom migration. On the contrary, at the volume diffusion, the solute
reached the following step can move further with the flow and the opposite
motion is always stable.

At large step velocities laying out of the frame of the quasi-static
approximation, the diffusion field of step has the power-type asymptotic
along the solution flow and the exponential decay in the opposite
direction (\ref{nonst}). Hence in this case, the step train is also stable
against the doubling of the period during motion in the opposite direction
with the flow and it is unstable during the parallel motion.

\section{Results and conclusions}
The linear stability analysis has shown that, at the opposite motion of
step and solution flow, the lateral instability is possible, and, at the
parallel motion, the instability against perturbation in the distances
between steps can occur. The overlap of the step diffusion fields
suppresses the lateral instability and gives rise the longitudinal
instability of step train. Intensification of the solution flow and
decrease of the step height stabilize the equidistant train of straight
steps. Though the instability has a diffusion nature, it can be even
noticeable near by the kinetic regime of the step motion. While the
quasi-static approximation is correct, the lateral instability becomes
stronger with an increase in the step velocity. Out of the quasi-static
frame, the effect of step motion plays a stabilizing role.

The evaluations have shown that the instability considered here can be
observed under the real conditions of solution growth of crystals like
KDP. As a rule at solution growth, meandering unit steps and presence of
the straight macrosteps allow to grow high quality crystals. A danger of
defect formation like solution inclusions does go from the presence of
meandered macrosteps. The concave segments of such steps further a
development of overhanging upper step layers, and then the formed cavity
can capture solution into the crystal. During the high-rate growth of
large KDP crystals, the crystal surface with macrosteps is commonplace, so
providing with uniformity of the diffusion layer thickness
\cite{Wilcox,EP} does not guarantee a stable growth without the inclusion
formation. Increasing the convection until the kinetic growth regime can
be also insufficient condition to obtain high quality crystals. It is
important to note that fast reversal of the flow direction can eliminate
the instability both lateral and longitudinal. If the interruption time
between equal intervals of the opposite solution flow is small in compare
with the time of the instability development $1/\nu$, then we find from
(\ref{nu}) that the lateral shape perturbations decays with the average
decay constant
\begin{equation}
\overline\nu=-\beta\lambda q^2 (1+\beta
U^{-1} I_0)^{-1}.
\label{mnu}
\end{equation}
A model of the diffusion layer relaxation when varying the flow velocity
was considered in Ref.  \cite{P}.

One of the goals for this study is the search of phenomena of the step
dynamics in order that the mechanism of the crystallizing matter transfer
should be determined. The distinct qualitative difference takes place for
step bunching. At the surface diffusion in the presence of the heating
current, step bunching appears both for step up and down direction of
the current, whereas at the volume diffusion in the solution flow, the
only steps moving parallel to the flow can be unstable.

The lateral instability considered here occurs both at the surface
diffusion due to the Schwoebel's effect and at the volume diffusion due to
the drift of the diffusion field by the solution flow. However, in the
case of the volume diffusion, the instability is controlled by the
intensity of the solution flow and the flow direction with respect to the
step motion. While interpreting experimental results, one should bear in
mind the possibility that a lateral instability can be excited by the
solution flow at the surface diffusion, because the delivery of the
crystallizing units to the crystal surface goes through the moving
solution at any rate. In the presence of the surface diffusion with the
diffusion length $\lambda_{2d}=(D_2\tau_s)^{1/2}$, where $D_2$ is the
surface diffusivity and $\tau_s$ is the characteristic time of the
desorption of the units back into the solution, the whole path of the
units includes diffusing in the solution to a sink strip with the width of
order of $\lambda_{2d}$ and then diffusing on the crystal surface within
this strip to the step. The drift of the volume diffusion layer by the
solution flow leads to a nonuniformity of the supersaturation along the
disturbed step, so the lateral instability is possible. In this case, only
long wave perturbations, $q<\tilde{A}/\lambda_{2d}$, where $\tilde{A}\sim
1$, can be unstable, corresponding critical wavelength depending on
neither the flow intensity nor the step velocity. Thus, these differences
of the step dynamics may be the criterion determining the delivery
mechanism.

At the same time with the diffusion processes, the morphological
instability can be conditioned by an impurity effect. The processes
resulting in the loss of vicinal-surface stability of the prismatic faces
of ADP and KDP under the action of the supersaturation nonuniformity and
the presence of growth-decelerating impurities had been observed by in
situ laser interferometry \cite{Rash}. It had been established that the
instability are most pronounced in the supersaturation range corresponding
to the maximum partial derivatives of the step velocity with respect to
the supersaturation and the impurity concentration. As it was found in Ref.
\cite{hyster}, an impurity influence can lead to the S-shape
dependence of the step velocity on the supersaturation and on the step
curvature too.  At the supersaturation near by this hysteretic range, the
step is unstable against the lateral perturbations. A model of instability
of the step spacings owing to impurity effect was considered in Ref.
\cite{Eerden}.  These instabilities conditioned by the impurities can be
separated through the dependence on the impurity concentration.

\acknowledgments
The author expresses his gratitude to Professor V.I.Bespalov for useful
discussion.
The research described in this publication was made possible in
part by Grant 95-02-03773-a from the Russian Foundation for Fundamental
Research and Grant JEM100 from the International Science Foundation and
Russian Goverment.

\section*{Figure captions}
Fig. 1. Scheme of singular crystal surface with single step.\\

\noindent
Fig. 2.
Integation path in expression (\protect\ref{z<0}).\\

\noindent
Fig. 3.
Scheme of the concentration isoline near by rectilinear step in the
solution flow.\\

\noindent
Fig. 4.
Dependence of $I_0$ on dimensionless step height $H=h/\Lambda$.\\

\noindent
Fig. 5.
Dependence of $\alpha(Q)$.\\

\noindent
Fig. 6.
Dependence of $I$ on dimensionless wave vector of lateral perturbation
$Q=q\Lambda$.\\

\noindent
Fig. 7.
Shaded zone corresponds to unstable lateral perturbation during contrary
motion of step and solution flow.\\

\end{document}